\documentclass{old_aa} % for the abstract without structuration 
\usepackage{graphicx}
\usepackage{txfonts}
\usepackage{natbib}

\usepackage{epsf}
\usepackage{rotating}
\usepackage{graphics}

\def\ms{\hbox{\,m\,s$^{-1}$}}         %m.s -1
\def\m2s2{\hbox{\,m$^{2}$\,s$^{-2}$}} %m2.s -2
\def\kms{\hbox{\,km\,s$^{-1}$}}       %km.s -1
\def \1s{$1\,\sigma$}

\def \t0{T$_0$}

\def \cible{HD\,80606}
\def \sophie{{\it SOPHIE}}

\begin{document}

   \title{Spin-orbit misalignment in the \cible\ planetary system\thanks{Based on observations 
   made with the 1.20-m and 1.93-m telescopes at Observatoire de Haute-Provence (CNRS), 
   France, by the \sophie\ consortium (program 07A.PNP.CONS), and 
   with a 16-inch telescope at Mt. Hopkins, Arizona, USA, by the MEarth team.}}
            
\author{Pont,~F. \inst{1}
\and H\'ebrard,~G. \inst{2}
\and Irwin,~J. M. \inst{3}
\and Bouchy,~F. \inst{2,4}
\and Moutou,~C. \inst{5}
\and Ehrenreich,~D. \inst{6}
\and Guillot,~T. \inst{7}
\and Aigrain,~S. \inst{1}
\and Bonfils,~X. \inst{6}
\and Berta,~Z.\inst{4}
\and Boisse,~I. \inst{2}
\and Burke,~C.\inst{10}
\and Charbonneau,~D. \inst{4}
\and Delfosse,~X. \inst{6}
\and Desort,~M. \inst{6}
\and Eggenberger,~A. \inst{6}  
\and Forveille,~T. \inst{6} 
\and Lagrange,~A.-M. \inst{6}
\and Lovis,~C. \inst{8}
\and Nutzman,~P.\inst{4}
\and Pepe,~F. \inst{8}
\and Perrier,~C. \inst{6} 
\and Queloz,~D. \inst{8}
\and Santos,~N.C. \inst{9}
\and S\'egransan,~D. \inst{8}
\and Udry,~S. \inst{8}
\and Vidal-Madjar,~A. \inst{2}
}

\institute{School of Physics, University of Exeter, Exeter, EX4 4QL, UK 
\and Institut d'Astrophysique de Paris, UMR7095 CNRS, Universit\'e Pierre \& Marie Curie, 98bis boulevard Arago, 75014 Paris, France
\and Harvard-Smithsonian Center for Astrophysics, 60 Garden Street, Cambridge, MA 02138, USA
\and Observatoire de Haute-Provence, 04870 Saint-Michel l'Observatoire, France
\and Laboratoire d'Astrophysique de Marseille, UMR 6110, CNRS\&Univ. de Provence, 38 rue Fr\'ed\'eric Joliot-Curie, 13388 Marseille Cedex 13, France
\and Laboratoire d'Astrophysique, Observatoire de Grenoble, Universit\'e J. Fourier, BP 53, 38041 Grenoble, Cedex 9, France
\and Universit\'e de Nice-Sophia Antipolis, Observatoire de la C\^ote d'Azur, CNRS UMR 6202, BP 4229, 06304 Nice Cedex 4, France
\and Observatoire de Gen\`eve, Universit\'e de Gen\`eve, 51 Chemin des Maillettes, 1290 Sauverny, Switzerland
\and Centro de Astrof\'isica, Universidade do Porto, Rua das Estrelas, 4150-762 Porto, Portugal
\and Space Telescope  Science Institute, 3700 San Martin Dr., Baltimore, MD21218, USA}

   \date{Received ; accepted }
      
  \abstract{We recently reported the photometric and spectroscopic detection 
of the primary transit of the 111-day-period, eccentric extra-solar planet \cible b, 
 at Observatoire de Haute-Provence, France. % (Moutou et al.~2009). 
The whole egress of the primary transit and a section of its central part were observed, 
allowing the measurement of the planetary radius, and evidence for a spin-orbit 
misalignment through the observation of the Rossiter-McLaughlin anomaly. The 
ingress having not been observed for this long-duration transit, uncertainties 
remained in the parameters of the system. We present here a refined, combined 
analysis of our photometric and spectroscopic data, together with further published radial 
velocities, ground-based photometry, and \textit{Spitzer} photometry around the secondary eclipse,
as well as new photometric measurements of \cible\ acquired at Mount Hopkins, Arizona, just before the beginning of the primary transit. 
Although the transit is not detected in those new data, they provide an upper limit for the transit duration,
which narrows down the possible behaviour of the Rossiter-McLaughlin anomaly in the unobserved part of the transit.
We analyse the whole data with a Bayesian approach using a Markov-chain Monte Carlo integration on all available information.
We find $R_{\rm p} = 0.98 \pm 0.03 \,{\rm R}_{\rm Jup}$ for the planetary 
radius, and a total primary transit duration of $11.9 \pm 1.3$ hours from first to fourth contact.
Our analysis reinforces the hypothesis of spin-orbit misalignment in this system (alignment excluded at $>95$ \% level), with a positive projected angle 
between the planetary orbital axis and the stellar rotation (median solution $\lambda \sim 50^o$). 
 As \cible\ is a component of a binary system, the peculiar 
orbit of its planet could result from a Kozai mechanism.}

%   \keywords{Planetary systems -- Techniques: radial velocities --  
% Techniques: photometry -- Stars: individual: HD\,80606}

  \authorrunning{Pont et al.}

   \maketitle

%________________________________________________________________

\section{Introduction}

\label{intro}

HD 80606 is a solar-type star with a gas giant planetary companion on a highly eccentric 111-day orbit \citep{nae01}. With $e=0.93$, the planet receives about a thousand times more star light at periastron than at apastron, which makes it a key system to study the atmospheric and thermal properties of hot gas giant planets. By a lucky coincidence (about 1 percent probability for a randomly oriented orbit), the orbital plane is aligned with the line-of-sight, so that both the secondary eclipse and primary transit  were detected. The secondary eclipse was measured during a long photometric run with the Spitzer space telescope  \citep{lau09}. In \citet[][hereafter M09]{mou09}, we presented our detection of the primary transit, simultaneously measured in photometry and spectroscopy with the 1.2-m and 1.93-m telescopes at Observatoire de Haute-Provence, France. The spectroscopic transit data seemed to indicate that the orbital plane of the planet was not aligned with the stellar rotation axis. But since the transit ingress was not observed, a large degree of uncertainty remained in this parameter, as well as in the latitude of the transit and radius of the host star.

Photometric data of the same event are available from two other teams
and locations. \citet[][F09]{fos09} presented data obtained at the Mill
Hill London Observatory, England, with two telescopes (a 35-cm Celestron
and a 25-cm Meade), on a time span that almost matches that of our OHP
observations the same night: data were obtained during the main portion
of the flat part of the transit, the whole egress, and a few hours
after its end. \citet[][G09]{gar09} presented data
obtained with a 60-cm telescope at the Esteve Duran Observatory, Spain,
on a shorter time span: the observations started just before the egress.
No detection of the transit ingress has been reported at the time of
writing.

In this paper, we present photometric data of \cible\ taken from Mt Hopkins, Arizona on the same night with the MEarth network \citep{irw09,nut08}. These data show no flux variation, but they do provide a powerful constraint on the system parameters by imposing a strict lower limit to the beginning of the transit.
We apply a Bayesian analysis to the whole data set, together with previous radial-velocity monitoring and the {\it Spitzer} observations near secondary eclipse, to calculate accurate values of the system parameters, including the radii of the host star and planet, and the spin-orbit angle.

\section{Observations}

\label{obs}

\subsection{MEarth photometry}

A single field containing HD~80606 and HD~80607 was monitored
continuously using one telescope of the MEarth observatory located at the Fred Lawrence
Whipple Observatory on Mount Hopkins, Arizona, for the night of
2009 February 13.  Additional observations were taken on 2009
February 14th and 15th but these are not used in the present work.
MEarth uses a non-standard $715\ {\rm nm}$ long-pass filter, with the
response limited at the red end by the long-wavelength tail of the CCD
quantum efficiency curve.

Observations were started at the end of nautical twilight (solar
elevation $12^\circ$ below the horizon), and continued until the start
of nautical twilight in the morning.  The airmass of the field varied
from $1.9$ at the start of observations, to a minimum of $1.1$ at
meridian transit, which occurred at UT 07:11, and at the end of
observations was $2.5$.  A total of $1695$ $4.8\ {\rm s}$ exposures
were obtained at an average cadence of $\sim 23\ {\rm s}$ including
overheads (CCD readout time, and re-centering the field after each
exposure).

Due to the extremely short exposures necessary to avoid saturation,
and the small telescope aperture of $0.4\ {\rm m}$, 
errors on individual measurements are dominated by atmospheric scintillation.  It was not
possible to defocus the telescope due to the small ($\sim 21''$)
separation between HD~80606 and HD~80607.   We use the formula
of  \citet{you67} to estimate the contribution to our observational
error bars arising from scintillation.  As noted by \citet{rya98},
 the typical coherence length for this effect is $\sim 12''$,
so the standard formula should remain a reasonable approximation.

Data were reduced using the standard MEarth reduction pipeline, which
is at present largely identical to the Monitor project pipeline
described in \citet{irw07}.  An aperture radius of $10\ {\rm
pixels}$ (corresponding to $7\farcs6$ on-sky) was used to extract
differential photometry.  In order to estimate the contamination in
the aperture for one star from the other given this large aperture
size, we use measurements in multiple concentric apertures from
single-stars in the same field as our target to derive a simple
curve-of-growth.  From this we estimate that $< 0.3\%$ of the flux
from HD~80607 falls inside the aperture centered on HD~80606, and vice
versa.  This is negligible for our purposes, since it would lead to a
$< 1\%$ underestimation in transit depths, which is smaller than
the observational error in this quantity.

%\begin{figure}
%\resizebox{8cm}{!}{\includegraphics{mearth_lc.pdf}}
%\caption{Photometric data from  MEarth on 13-14 Feb 2009.}
%\label{mearth_fig}
%\end{figure}

We used HD~80607 as a comparison star to derive differential light
curves of HD~80606 from this photometry.  Since these stars have 
similar positions on the detector and almost identical colors, effects
such as color-dependent atmospheric scintillation and flat fielding
error are minimized by doing this, and we find no advantage to
attempting to use other stars of comparable brightness on the field as
additional comparison stars.  MEarth uses German Equatorial Mounts, so
the entire telescope and detector system must be rotated through
$180^\circ$ relative to the sky upon crossing the meridian.  Using
HD~80607 as a comparison star, we see little evidence for flat
fielding errors in the data for HD~80606, which normally manifest as
different base-line levels in the light curve for positive and
negative hour angle.  We therefore apply no correction for this effect
in the present analysis. 

The MEarth data is given in Table~\ref{mearth_tab}. %The binned time series for February 13 is shown on Fig.~\ref{tr_fig}.

\begin{table}
\centering
\begin{tabular}{lll} 
\hline
Date [HJD] & F$_{715}$ [mag] & $\sigma_F$ \\ \hline 
2454875.589851 & 8.116900 & 0.005537 \\
2454875.590117 & 8.133620 & 0.005522 \\
2454875.590395 & 8.142982 & 0.005507 \\
2454875.590684 & 8.160600 & 0.005493 \\
2454875.590950 & 8.126864 & 0.005480 \\
2454875.591217 & 8.142918 & 0.005471 \\
2454875.591494 & 8.129089 & 0.005450 \\
2454875.591749 & 8.153709 & 0.005435 \\
2454875.592015 & 8.074484 & 0.005425 \\
... & .. &  ... \\ \hline
\end{tabular}
\caption{Photometric times series for HD 80606 from MEarth (full table available electronically)}
\label{mearth_tab}
\end{table}

\subsection{OHP 120-cm photometry}

The photometric observations of \cible\ and HD\,80607 performed at the
120-cm telescope at OHP during the nights 2009 February 12 and
13 were presented by M09. The transit was detected during the second
night, which is the only one of the two that we use in the present work.
326 frames were secured during the transit night, with 20 to 30-second
exposure times. The negative slope after egress tends to suggest that
a correction for airmass variations should be taken into account in
deriving the flux. The airmass is 1.5 at the beginning of the night,
then reaches 1.0 at the middle of the egress, and increases up to 1.66
at the end of the night. Assuming that the slope seen
out of transit is due to airmass changes, and that the effect of
airmass on the photometry is linear,
we find a correction of $(2-{\rm airmass})\times 0.004$.
The impact of the correction is to create a small slope at the beginning
of the transit sequence, i.e. during the flat section of the transit. The
significance of this new slope is low. Since this correction 
dominates the error budget on the transit parameters, it is important to
take correlated noise into account in the analysis.

\subsection{\sophie\ Doppler spectroscopy}

Radial velocities of \cible\ measured with the \sophie\ spectrograph at the 1.93-m telescope of OHP 
were presented by M09. A continuous sequence of 39 measurements 
was acquired during the transit night (2009, $13-14$ February), as well as nine extra 
measurements between $8\mathrm{th}$ and $18\mathrm{th}$ February, out of the transit.
The radial velocities were obtained from a weighted cross-correlation of the spectra with a G2-type 
numerical template. All the spectra present 
a similar signal-to-noise ratio, S/N~$\simeq 47$ per pixel at 550~nm. Together with the parameters 
of the cross-correlation function of the spectra (full width at half maximum of $7.33 \pm 0.02$~\kms, 
and a contrast representing $48.2 \pm 0.3$~\%\ of the continuum), this corresponds to a $\sim2.4$~\ms\ 
photon-noise uncertainty on the radial velocity measurements. We quadratically 
added 3.5~\ms\ due to telescope guiding errors and 1~\ms\ due to wavelength calibration, resulting 
in radial velocities with a typical accuracy of $\sim4.5$~\ms.

\subsection{Published Doppler and photometric data}

We complemented our \sophie\ measurements with published radial velocities from three 
other instruments: 74 measurements from \textit{ELODIE} at OHP \citep{nae01,mou09}, 46 \textit{HIRES} measurements from Keck
\citep{but06}, and 23 \textit{HRS} measurements from the Hobby-Eberly 
Telescope \citep{wit09}. 
The typical accuracies of these three extra datasets are 12.3,   5.0, and  8.3~\ms\  
respectively, and their time spans are Nov.~1999--Nov.~2003, Apr.~2001--Feb.~2005, 
and Dec.~2004--Mar.~2007, respectively. None of those measurements were obtained during 
a primary transit.
These measurements are used to refined the Keplerian orbit of the planet. Systematic 
radial velocity shifts between the different datasets are unknown, and left as free parameters. The residuals
around the model orbit show that the  \textit{ELODIE}  uncertainties are underestimated, and we scaled these uncertainties
upwards by a factor 1.8 in order to obtain coherent normalized residuals.
%In a similar manner, the \sophie\ measurements performed out of the transit constraint the center-of-mass 
%velocity of these measurements with respect to the three other datasets.

The egress of the transit of 14th February 2009 was observed in photometry from three locations covering part of the transit center and the transit egress (M09, F09, G09; see Section~\ref{intro}).

\cible\ was also monitored for nearly 24 hours around the time of the secondary eclipse at 8 microns with the Spitzer space telescope \citep{lau09}. These data provide strong additional constraints on the system parameters by measuring the time and duration of the secondary eclipse.

We include these data sets in our combined solution (we use only the Celestron time series from F09).

\section{Analysis}

\begin{figure}
\resizebox{10cm}{!}{\includegraphics{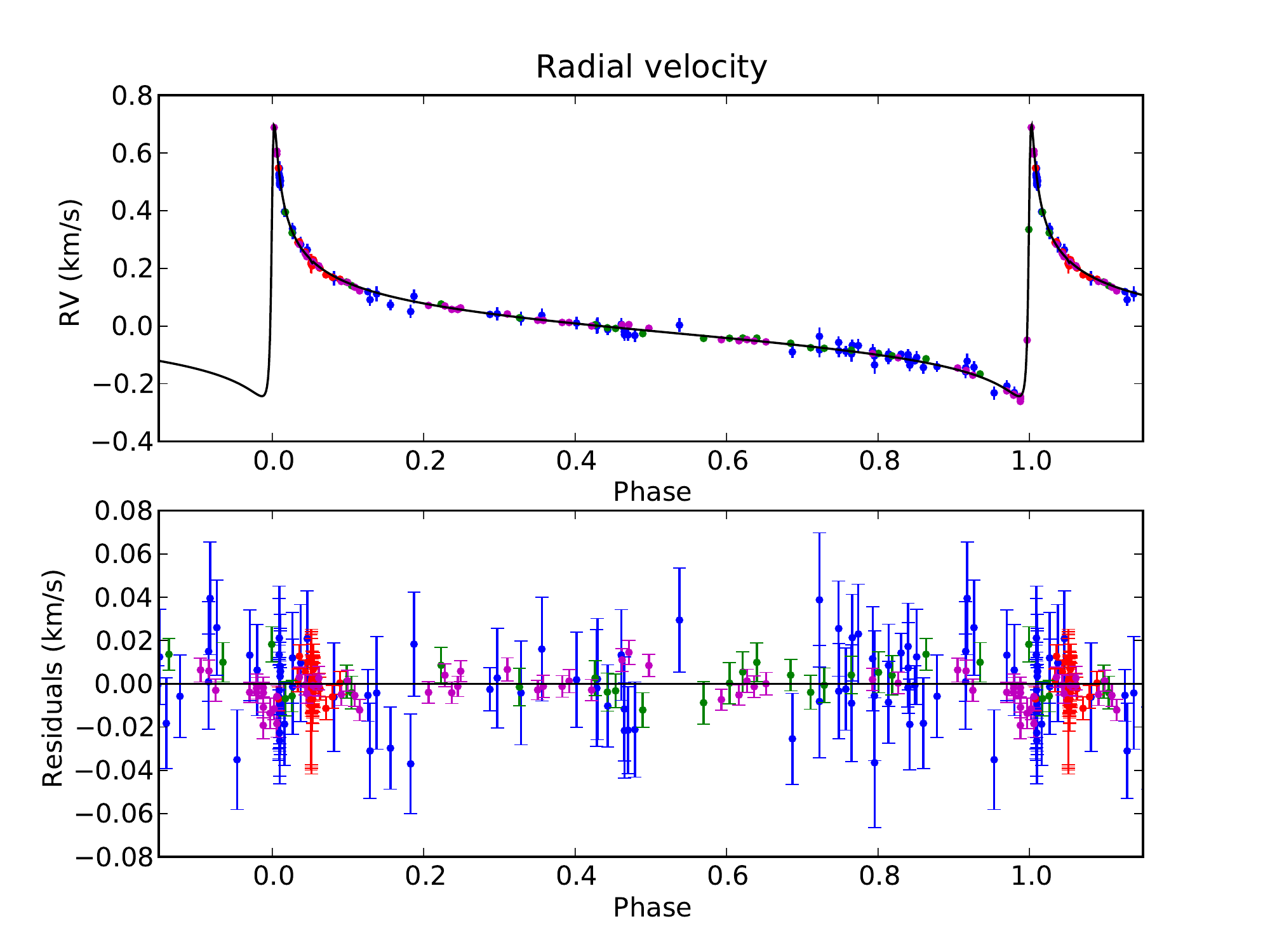}}
\caption{Radial-velocity data and model. Data from Elodie (blue), Sophie (red), Keck (magenta) and HET (green).}
\label{vr_fig}
\end{figure}

\begin{figure*}
\resizebox{16cm}{!}{\includegraphics{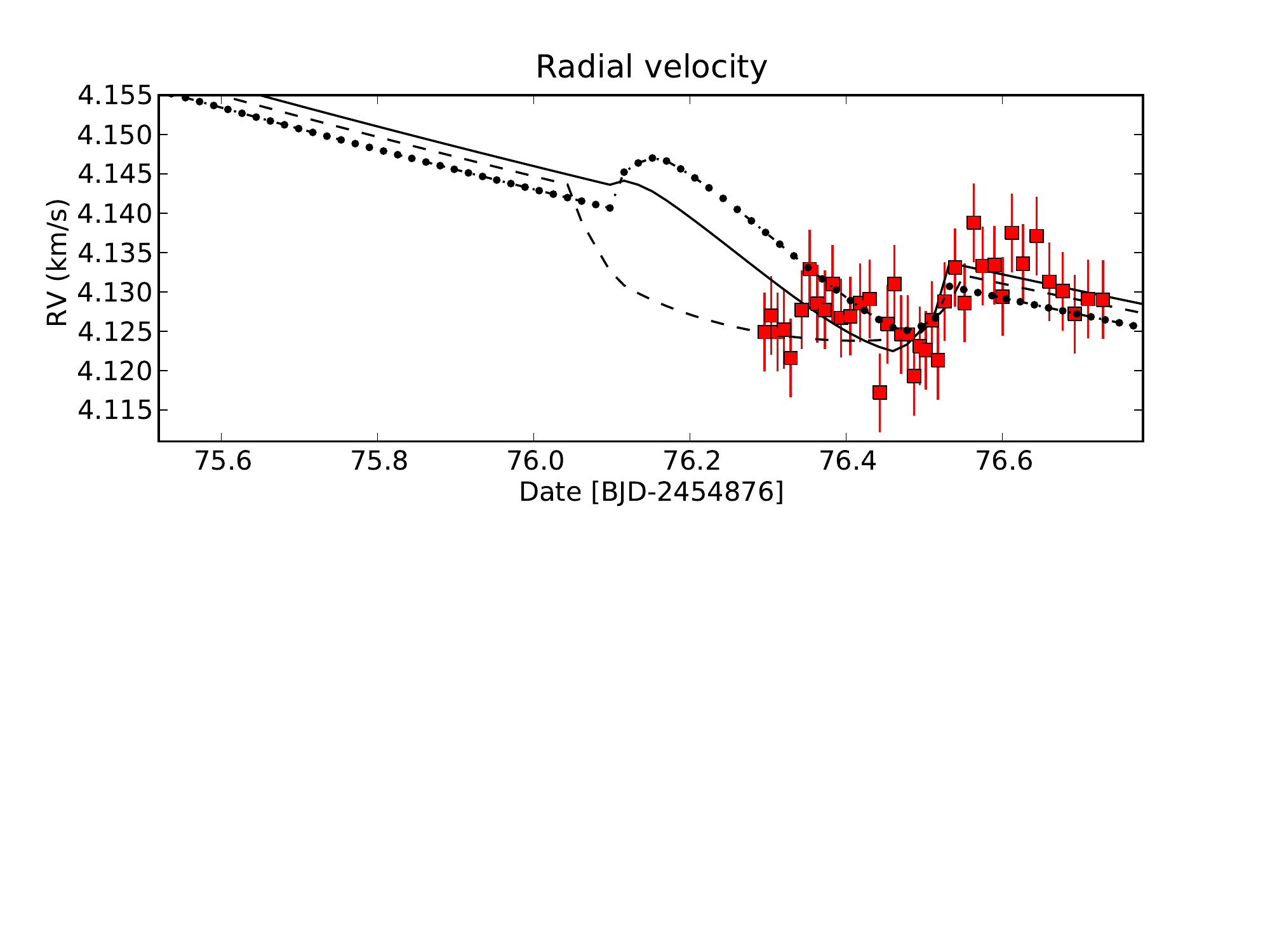}}
\vspace{-5.5cm}\caption{Radial-velocity data and models around the phase of transit. The three model curves correspond to spin-orbit angles at the best-fit solution ($ \lambda=33^0$, solid line), a good solution with a higher spin-orbit angle ($\lambda=122^0$, dashed), and the highest-likelihood solution with aligned spin and orbit ($\lambda=0$, dash-dotted).}
\label{rm_fig}
\end{figure*}

\begin{figure*}
\resizebox{16cm}{!}{\includegraphics{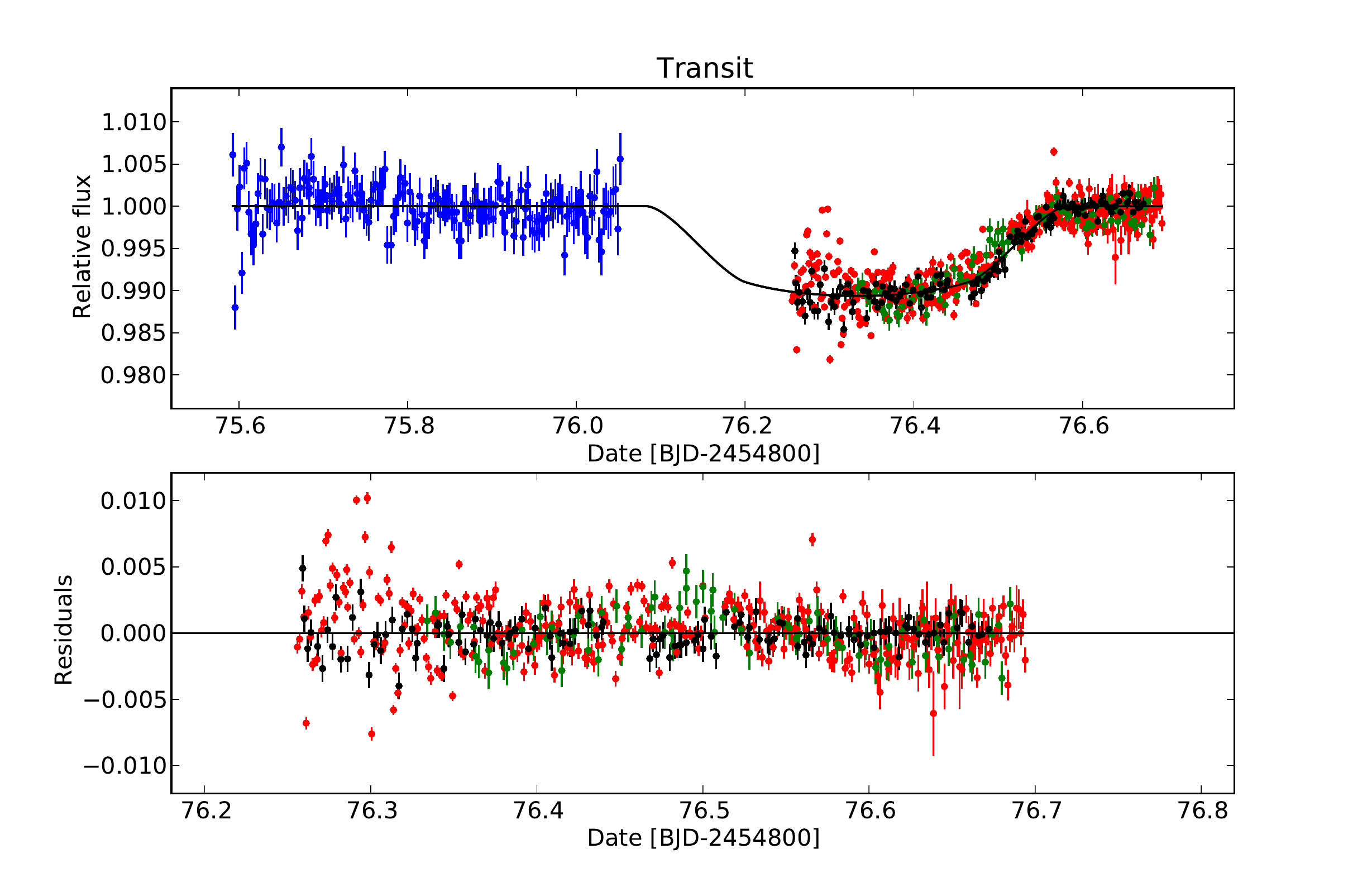}}
\caption{Photometric data on 13-14 Feb 2009 transit. Red: M09, black: MEarth, blue: F09, green: G09 }
\label{tr_fig}
\end{figure*}

Figures~\ref{vr_fig},~\ref{rm_fig} and ~\ref{tr_fig} display the photometric and spectroscopic data for the radial velocity curve, the spectroscopic transit and the photometric transit. The secondary eclipse is plotted in \citet{lau09}. Figure~\ref{geom} shows the configuration of the system according to the best-fit solution in Section~\ref{results}.

The procedures to infer physical parameters from observations of transiting planetary systems have now been firmly established, and numerous descriptions can be found in the recent literature \citep[see e.g. contributions to][for reviews and examples]{iau253}. The case of HD 80606b requires all the toolbox of the trade for several reasons: the extreme eccentricity makes the relation between physical parameters and observable quantities even more non-linear than usual, the incomplete coverage of the transit in photometry and spectroscopy requires sound Bayesian statistics, most data sets are dominated by correlated noise rather than random errors, and different types of information (transit and secondary eclipse photometry, radial velocity orbit, stellar evolution models) provide partial constraints of comparable importance. For these reasons, we need to use a fully Bayesian approach, with physically meaningful priors on all parameters. We also need proper accounting for correlated noise.

%\begin{figure}
%\resizebox{8cm}{!}{\includegraphics{sec_fig.pdf}}
%\caption{{\it Spitzer} photometric data around the secondary transit.}
%\label{sec_fig}
%\end{figure}

Although the data is incomplete in some respects (absence of detection of the transit ingress), the abundant datasets leave little leeway for most parameters. In particular, the combination of the secondary eclipse duration, the egress duration and the orbital parameters tightly constrain the mass and size of the host star and the planet.

\subsection{Method}

We use a Bayesian approach to determine the probability distribution of the physical parameters given all the available observations and prior information such as stellar evolution models.  We perform the integration with a Markov Chain Monte Carlo (MCMC) algorithm, using Metropolis-Hastings sampling. This is similar to the procedure described for instance in \citet{cam07}. 

This technique is particularly sensitive to the assumptions used on the distribution of measurement uncertainties. MCMC integration samples the merit function used to measure the agreement between model and observations according to the Bayesian prior probability, and as such it cannot be better than the merit function used. If a traditional ``observed vs. measured'' sum-of-squares is chosen, the error underestimation due to correlated noise that plague straightforward fitting methods will also be observed. To avoid biased results and unrealistically narrow posterior probability distributions for the output parameters, it is essential to use realistic error estimates for all data, including systematic errors. In the photometric and transit spectroscopic time series of \cible\, the dominant source of systematics are fluctuations correlated in time (e.g. airmass and seeing effects). \citet{pon06} discuss the effects of this ``red noise'' and a possible way to integrate it in the merit function.

\subsection{Correlated noise}

For most datasets pertaining to HD 80606, red (i.e. correlated) noise is by far the dominant source of uncertainties: (1) the \sophie\ coverage of the spectroscopic transit was acquired during a single night. It is known that in such conditions, \sophie\ data are sensitive to weather-related systematics at the level of a few meters-per-second; (2) the photometric times series also cover only the transit egress. When only a partial transit is observed with ground-based photometry, the correction of slow trends is difficult  and dominates the error budget; (3) the out-of-transit baseline flux of the {\it Spitzer} observations is made variable at the $10^{-4}$ level by instrumental effects and by the light reflected on the day side of the planet, a factor which, from the point of view of parameter determination, is equivalent to red noise.

We model the red noise as described in Pont et al. (2006), with a single $\sigma_r$ parameters for each data set, describing the amplitude of correlated noise over the relevant timescale. For the photometric time series we use the ``$V(n)$'' method described in that paper, which estimates $\sigma_r$ from the departure of binned versions of the data compared to purely white noise. We find $\sigma_r=7\times 10^{-4}$ for the OHP photometry, $8\times 10^{-4}$ for the MEarth photometry, $3.6\times10^{-4}$ for the F09 photometry , $8\times 10^{-4}$ for the G09 photometry, 4 $m\, s^{-1}$ for the Sophie transit spectroscopy and $2\times 10^{-4}$ for the {\it Spitzer} time series.
The first three values are coherent with our experience of ground-based transit spectroscopy, with $\sigma_r$ between 4 and $10 \times 10^{-4}$ being typical for single-object rapid cadence time series. The value of the correlation parameter for the \sophie\ time series is also compatible with previous experience  \citep[e.g.][]{heb09}. Finally, there is enough out-of-transit data in the {\it Spitzer} time series to measure $\sigma_r$ precisely. 
Our red noise analysis thus indicates that the F09 Celestron data is subject to lower systematics than the M09 and G09 photometric measurements (as visible as well in the behaviour of the residuals compared to model lightcurves). This implies that the first data set will get about twice as much weight as each of the other two in the combined analysis.
To account for the clearly much lower quality of the photometric data at the beginning and end of the night, we increase $\sigma_r$ by a factor three for these measurements ($JD<2454876.34$).

\subsection{Models}

We use the \citet{man02} algorithm to build model transit and secondary eclipse lightcurves, with a linear limb-darkening law with $u=0.66$ (suitable for a solar-type star; at this point the accuracy of the data does not require higher-order modelling of the limb darkening). We use the \citet{oht05} analytical description of the RM effect, and a Keplerian radial-velocity orbit. We compute the 3-D position of the planet relative to the star at each date. Two additional parameters are introduced to describe the photometric signal of the secondary eclipse: a planet-to-star surface-brightness ratio in the Spitzer band, and an out-of-transit continuum flux. The  brightness variation of the planet around the time of secondary eclipse is neglected. 

\subsection{MCMC}

We use a chain with 200000 steps, starting with the parameters in M09. The size of the steps is set to 0.05 times the uncertainties quoted by that study. The free parameters are the following: $P, T_0, i, e, \varpi, V_0^{1..4}, M_{\rm pl}, R_{\rm pl}, M_{\rm s}, R_{\rm s},  dT, \lambda, V_{\rm rot}$, respectively the period, epoch of periastron, orbital inclination, orbital eccentricity, argument of periastron, center-of-mass radial velocity (for each spectrograph independently), planet mass and radius, star mass and radius, planet-to-star surface brightness ratio in the Spitzer band,  spin-orbit angle and projected stellar spin. We use Gibbs sampling for the steps in the chain (next step accepted with a probability equal to the ratio of likelihoods).  We include the constraints from stellar evolution models directly  in the prior and in the merit function used in the MCMC: the MCMC chain moves in an interpolation of the \citet{gir02} stellar evolution models with uniform steps in mass, age and metallicity (corresponding to a flat prior in these quantities). Each evolution model produces values for the stellar mass, radius and temperature, $M_s$, $R_s$ and $T_{\rm eff}$, that are used to estimate the merit function. How this method solves Bayes'  theorem is studied in more details in the context of stellar ages in \citet{pon04}.

\subsection{Merit functions}

The merit function in the MCMC is taken as the likelihood of the observations given the model, assuming gaussian error distributions and purely ``white+red'' noise \citep{pon06}, using all the photometric and radial velocity data, as well as the spectroscopically determined stellar parameters:

\begin{eqnarray}
\chi^2 = \sum_{\rm phot} \frac{(F_{\rm obs}-F_{\rm mod})^2}{\sigma_{Fw}^2+N\sigma_{Fr}^2} + \sum_{\rm rv} \frac{(V_{\rm obs}-V_{\rm mod})^2}{\sigma_{Vw}^2+N \sigma_{Vr}^2} \\
+ \sum_{\rm star} \frac{(S_{\rm obs}-S_{\rm mod})^2}{\sigma_S^2}
\end{eqnarray}

where $F$ are the photometric data, $V$ the radial velocity data, $S$ the stellar parameters, and the {\it obs} and {\it mod} subscripts denote observed and model values respectively. The $w$ and $r$ subscript indicate  white and red noise parameters, and $N$ is the number of data points during the correlation timescale. $F_{\rm mod}$ is the model light curve for the transit and secondary eclipse, $V_{\rm mod}$ the model radial velocity curve including the RM anomaly during transit, and $S_{\rm mod}$ the set of observable parameters of the host star according to the \citet{gir02} models. The sums are made respectively on the individual photometric observations, spectroscopic observations, and input stellar parameters of the evolution models (age, mass and metallicity). We take the duration of the egress as the relevant correlation timescale to evaluate $N$ ($\sim$ 25 points for photometric time series, 5 points for the radial velocity sequence).

\subsection{Prior distributions}

We define the steps in the MCMC corresponding to flat prior distributions in the following quantities: $P$, $T_0$, $\cos i$, $V_i$, $e \cos \varpi$,  $e \sin \varpi$,  $K\sqrt{1-e^2}$, $R_{p}/R_{\rm s}$,  $M_s$, $\tau_{\rm s}$,  [Fe/H], $dT$, $\cos \lambda$. The combinations were chosen either for physical reason (isotropic orientation of spin and orbit in space, random epoch), to avoid strong covariance in the MCMC (radius ratio instead of radius, eccentricity dependence of $K$ integrated in the prior) or linear when the prior is not significant compared to the observational constraints ($P, V_i$).  

The prior distribution in the parameters of the parent star ($M_s, R_s, T_{\rm eff}$) corresponds to a flat distribution in age, mass and metallicity, as described above.  We therefore do not reduce the information from stellar evolution models to a single term in the merit function, but make use of the full information provided by the stellar evolution models, thus taking into account the correlation of stellar mass, radius and temperature, and the respective probability of different models for a field star in the solar-neighbourhood. 

We set the stellar rotational velocity to $V_{rot}=1.8$ km$\,$s$^{-1}$ \citep{fis05}. This parameter influences the shape of the spectroscopic transit radial velocity curve, but is poorly constrained by the current data. The results do not vary significantly if we repeat the procedure with a moderate uncertainty on this value (up to about 1 km$\,$s$^{-1}$). If this parameter is left completely free, the Markov chain does not converge, because different combinations of the rotation velocity, spin-orbit angle and impact parameter produce similar predictions for the observed portion of the spectroscopic transit.

Altogether, three prior distributions  have a significant effect on the solution:  the orbital inclination angle (penalizing grazing transits compared to a non-Bayesian fit), the stellar age prior (penalizing rare or inexistant combinations of stellar parameters), and the stellar rotation velocity (lifting the main degeneracy in the spectroscopic transit).

\section{Results and discussion}

\label{results}

\begin{figure}
\resizebox{8cm}{!}{\includegraphics{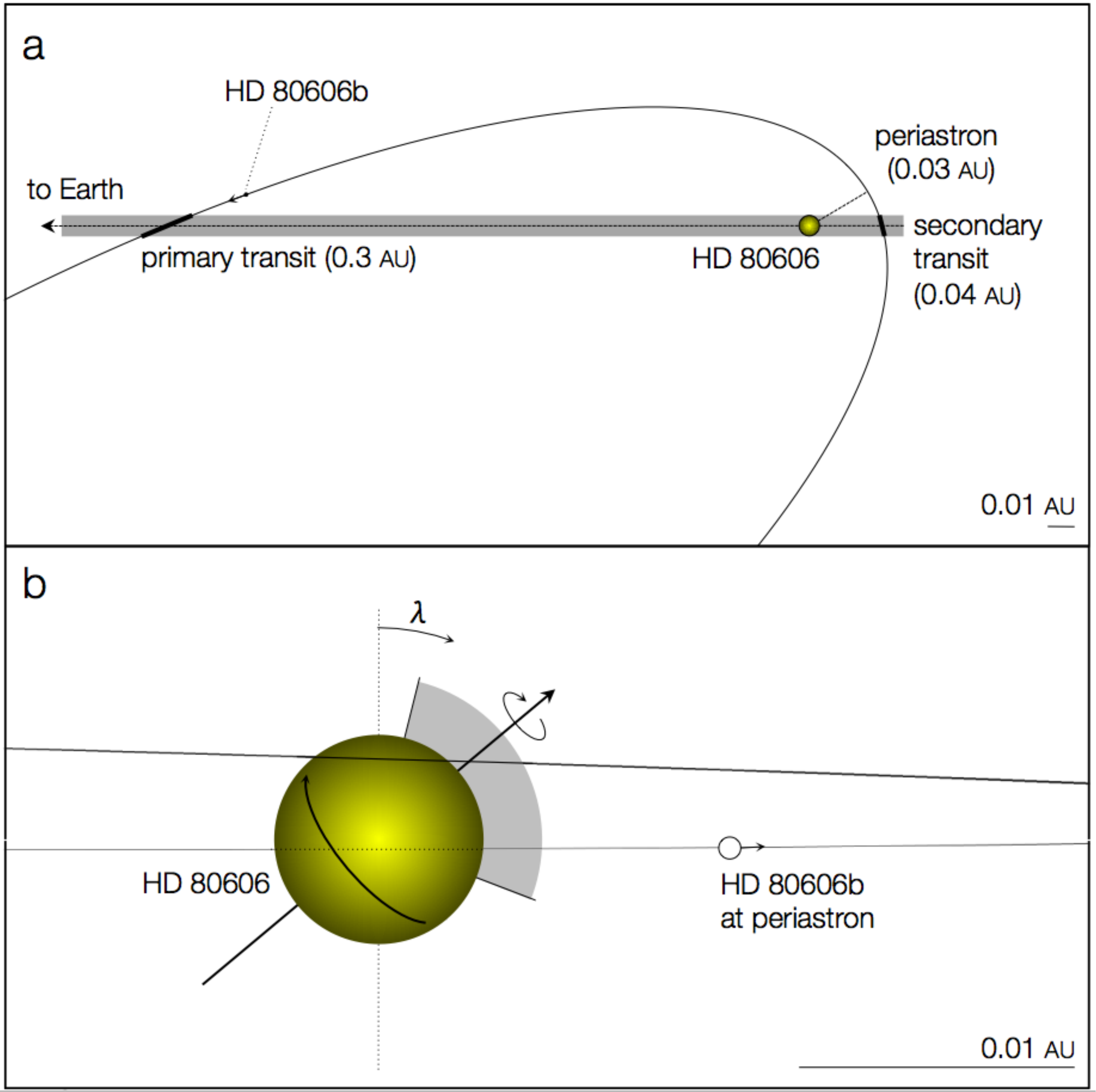}}
\caption{Geometry of the HD 80606 system according to our best-fit solution, {\bf (a)} from above the orbit, {\bf (b)} seen from Earth.}
\label{geom}
\end{figure}

Table~\ref{param} gives the median values and central 68\% confidence intervals for the parameters of the HD80606 system given by the MCMC integration. Note that because the median values correspond to different individual solutions for each parameter, the orbital parameters in the table do not correspond to the best-fit orbit. The best-fit orbit is $P=111.43605$ days, $K=476.48$ m$\,$s$^{-1}$, e=0.9332, $T_0=2454424.8529$ BJD, $\varpi=300.71^0$.

\begin{figure}
\resizebox{8cm}{!}{\includegraphics{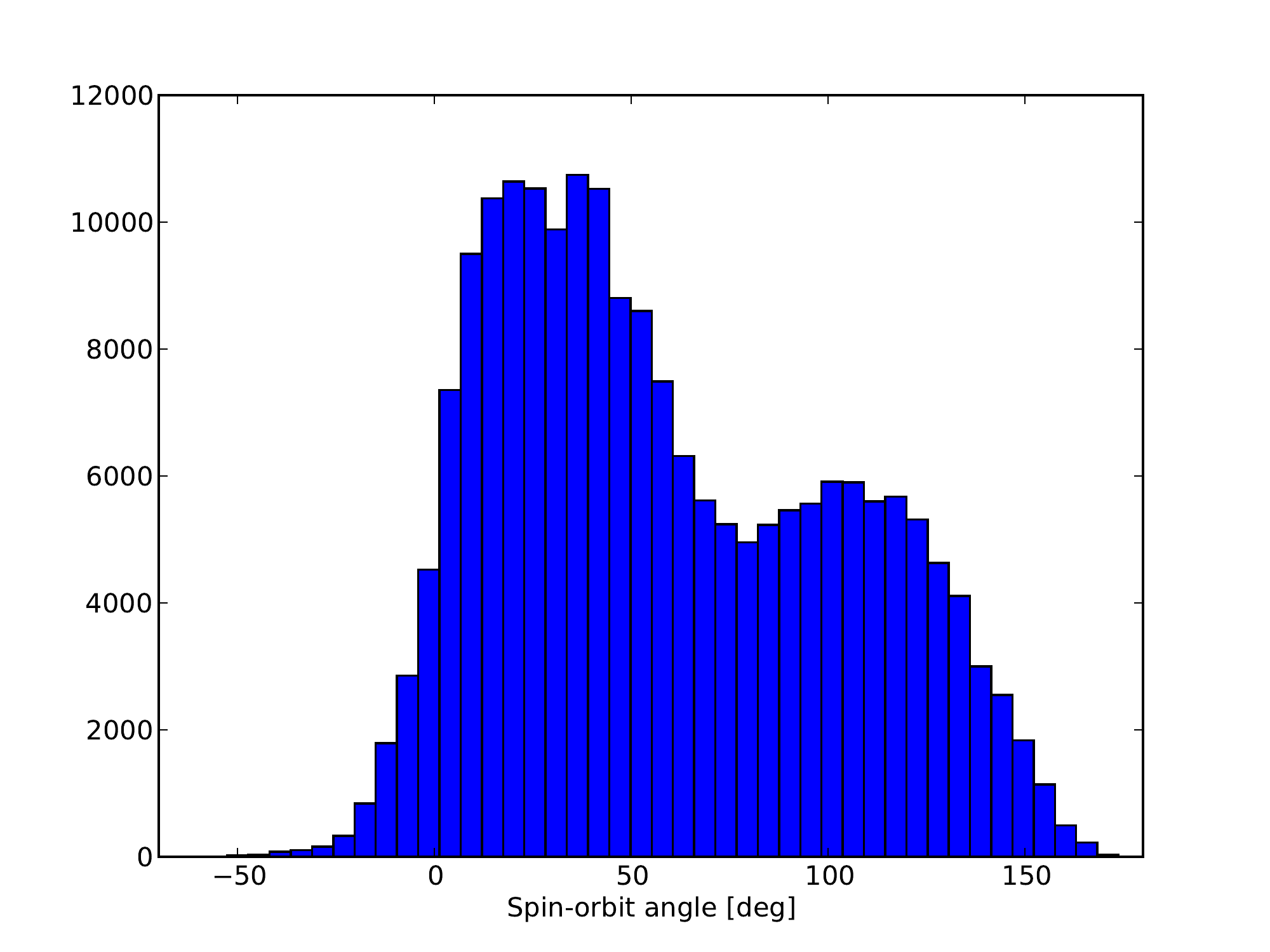}}
\caption{Posterior probability distribution for the projected spin-orbit angle.}
\label{lam}
\end{figure}

\begin{figure}
\resizebox{8cm}{!}{\includegraphics{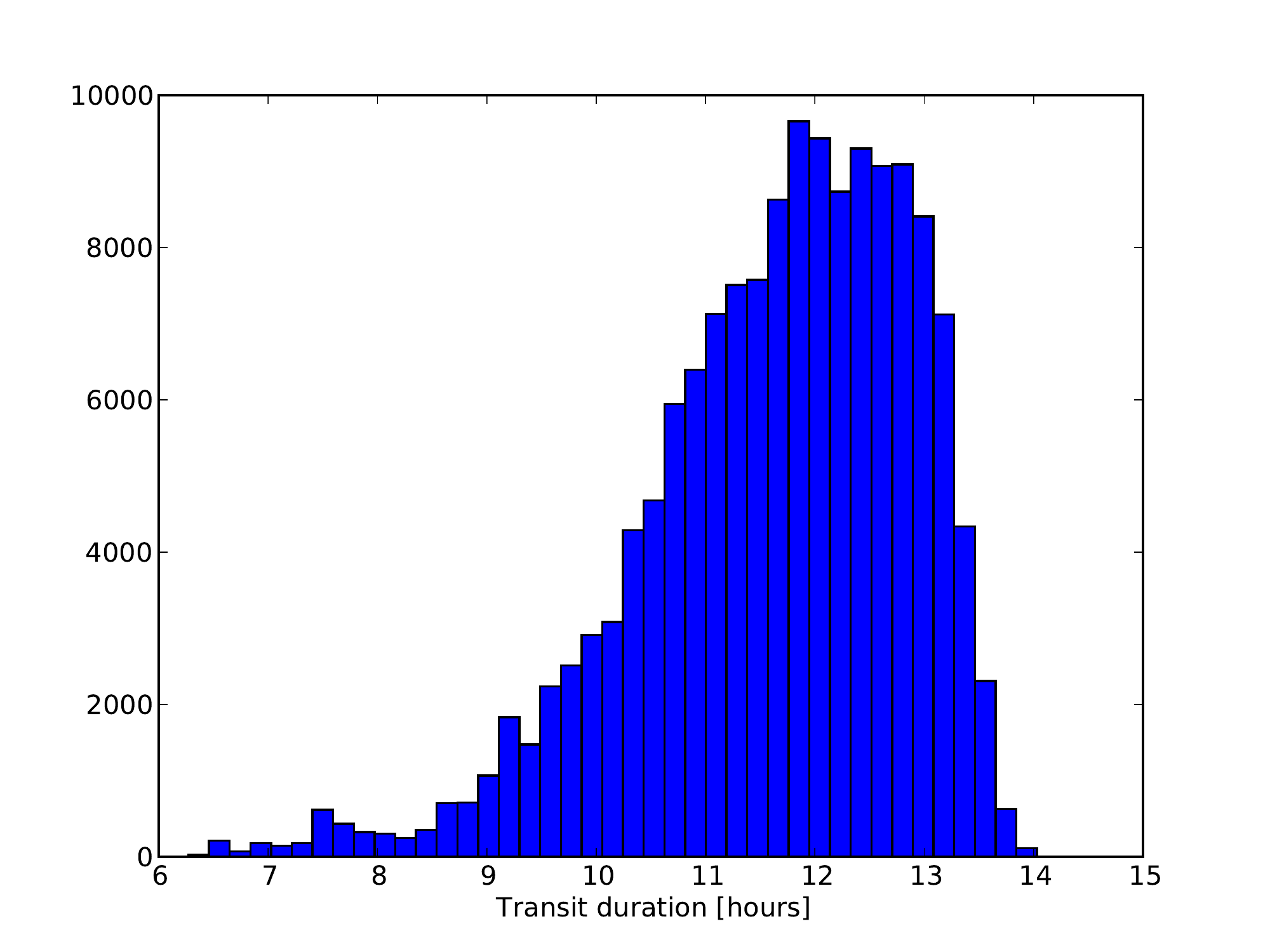}}
\caption{Posterior probability distribution for the transit duration (first to fourth contact)}
\label{Ttr}
\end{figure}

Figures~\ref{lam} and \ref{Ttr} show the probability distribution functions for the spin-orbit angle and transit duration (first to fourth contact).

\begin{table}
\centering
\begin{tabular}{l  r }
\hline
Center-of-mass velocity $V_0$ (Elodie) &3.787 $\pm$ 0.004 km$\,$s$^{-1}$ \\ 
\hspace{3cm} $V_0$(SOPHIE)& 3.910 $\pm$ 0.004 km$\,$s$^{-1}$\\ 
\hspace{3cm} $\Delta V_0$(Keck)& $-0.001 \pm 0.002$ km$\,$s$^{-1}$\\ 
\hspace{3cm} $\Delta V_0$(HET)& $-0.019 \pm 0.003$ km$\,$s$^{-1}$\\ 
Orbital period $P$ &111.4357 $\pm$ 0.0008 days \\
Orbital eccentricity $e$ &0.9332 $\pm$ 0.0008  \\
Velocity semi-amplitude $K$ &474 $\pm$ 4 m$\,$s$^{-1}$\\ 
Epoch of periastron $T_0$ & 2 454 424.852 $\pm$ 0.008 BJD\\ 
Argument of periastron $\varpi$ & 300.80 $\pm$ 0.22 $^o$\\ 
Orbital inclination $i$  & 89.32 $\pm$ 0.06 $^o$\\
Semi-major axis $a$ & 0.449 $\pm$ 0.006 AU \\
& \\
Epoch of transit $T_{\rm tr}$ & 2454876.316 $\pm$ 0.023 BJD \\ 
Epoch of eclipse $T_{\rm e}$& 2454424.719 $\pm$ 0.009 BJD \\ 
Transit duration  $T_{1-4} $& 11.9 $\pm$ 1.3 hours \\
Transit duration  $T_{2-3} $& $9.6^{-1.3}_{+0.8}$ hours \\
Radius ratio $R_p /R_s$ &0.103 $\pm$ 0.003 \\ 
Spin-orbit alignment $\lambda$ &$ 50 ^o$ [ 14 -- 111 ] $^o$  \\
Impact parameter $b$ & 0.75 $\pm$ 0.06 \\
& \\
Star Mass $M_s$& 0.97 $\pm$ 0.04 M$_\odot$ \\
Star Radius $R_s$ &0.978 $\pm$ 0.015 R$_\odot$ \\
%Projected spin $V  \sin I_s$ & 1.2 $\pm$ 0.4 km$\,$s$^{-1}$ \\
& \\
Planet mass $M_p$ &3.94 $\pm$ 0.11 $M_J$ \\
Planet radius $R_p$  & 0.98 $\pm$ 0.03 $R_J$ \\ \hline
\end{tabular}
\caption{Parameters for the HD 80606 system. Uncertainties from the 68\% central probability interval of the posterior distribution traced by the Markov chain.}
\label{param}
\end{table}

\begin{figure*}
\resizebox{16cm}{!}{\includegraphics{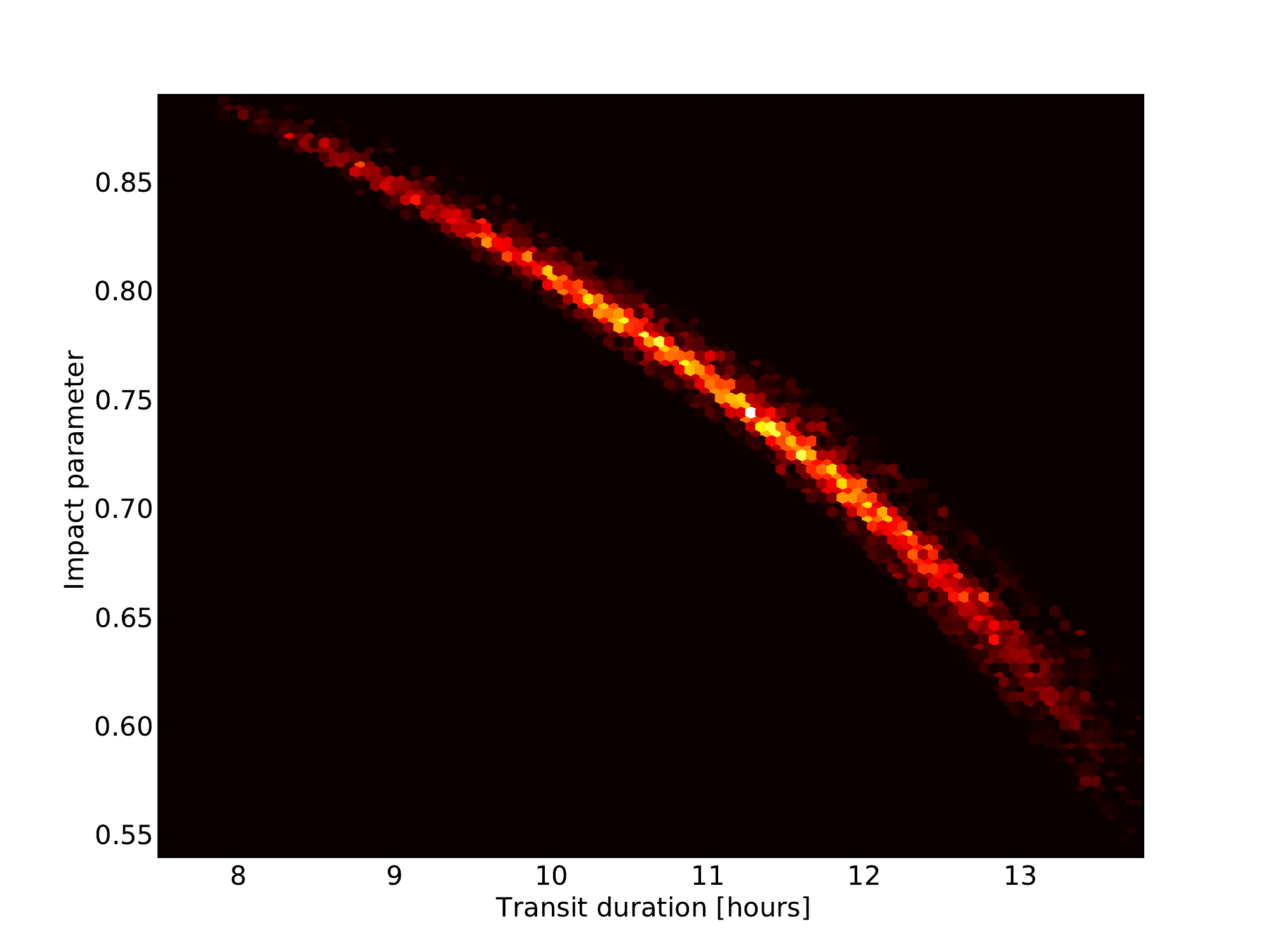}\includegraphics{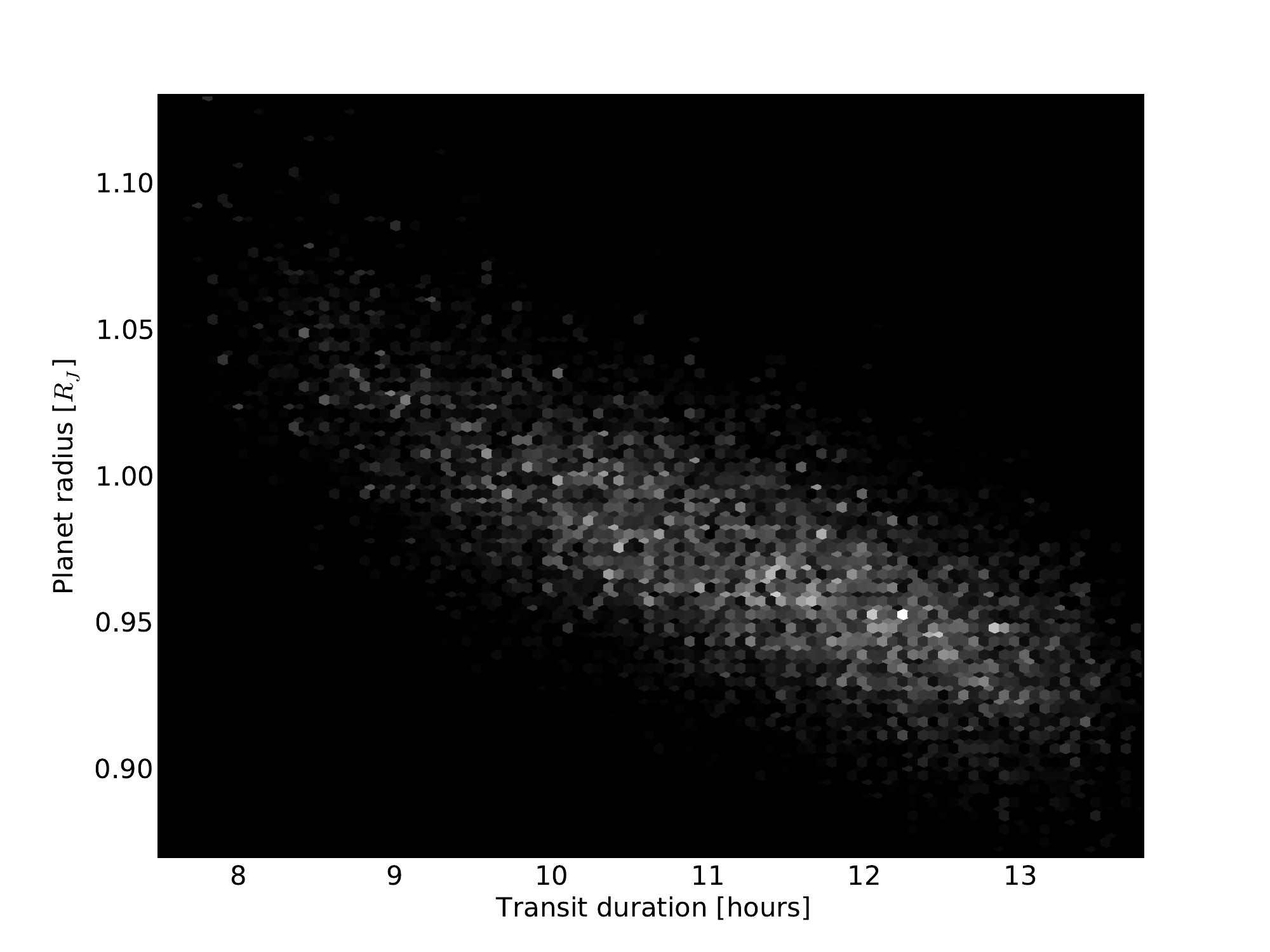}}
\caption{Posterior probability density: transit duration vs impact parameter (left) and transit duration vs planetary radius (right)}
\label{mc_fig}
\end{figure*}

The strongest covariance between the output parameters is that of the impact parameter with the transit duration. In general, the transit duration depends mainly on the radius of the host star and impact parameter. When the whole transit is measured, the main covariance is between impact parameter and host star radius.  However, in our case, the star radius is relatively well constrained by the duration of the secondary eclipse, and the transit duration is weakly constrained because the transit ingress was not observed, so that the main covariance is between the transit duration and impact parameter (higher impact parameters corresponding to shorter transit).
Fig.~\ref{mc_fig} shows the posterior probability density in the $D$ vs. $b$ plane, and in the $D$ vs $R_p$ plane. The core of the probability distribution in each parameter is well described by Normal distributions, with a tail of low-probability solutions at high impact parameter, low transit duration and larger planet.  

\begin{figure}
\resizebox{8cm}{!}{\includegraphics{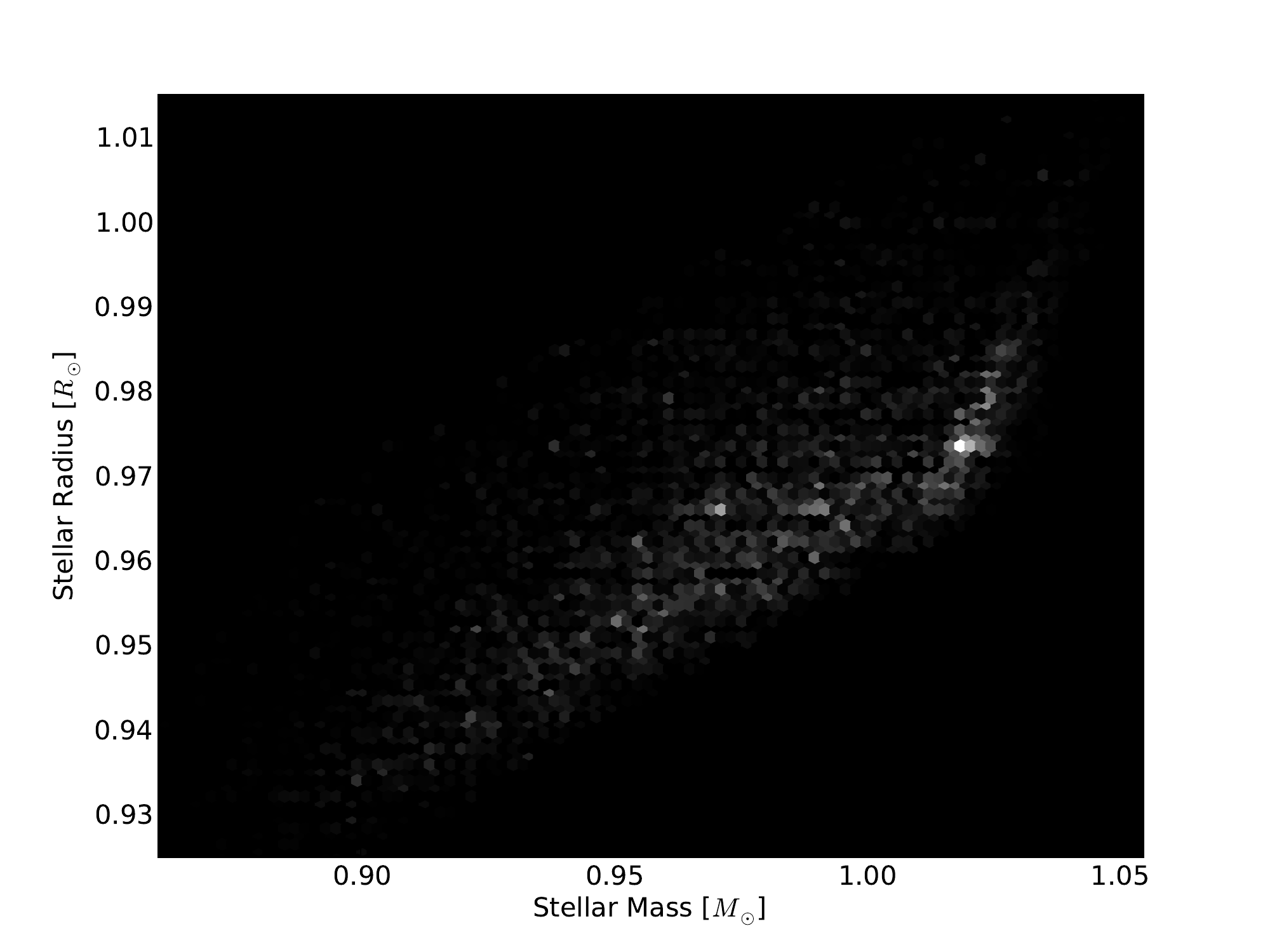}}
\caption{Posterior probability density: stellar mass and radius.}
\label{MR_fig}
\end{figure}

Figure~\ref{MR_fig} shows the probability density in stellar mass and radius. The main correlation is $M\sim R^{3}$, because the transit parameters are degenerate in these quantities. The sharper limit towards larger masses is due to stellar evolution models.

%The results depend on the set of stellar evolution models used. We have compared the predictions of the \citet{gir02} stellar evolution models with the CESAM models (REF). We find that for spectroscopic parameter like those of \cible, the CESAM models predict masses about 10\% larger, which would correspond to radii $\sim 3$ \% smaller for the star and planet. {\bf Elaborate. How is such a large difference possible so near to Solar values?}

\subsection{Uncertainty intervals}

F09 quote very small uncertainties for the transit duration, orbital inclinations and planetary radius (1.029 $\pm$ 0.017 $R_J$, 12.1 $\pm$ 0.4 hours, $89.285 \pm 0.023$ degrees). The uncertainties are small because  F09 keep several important parameters fixed (the host star radius and  orbital parameters), and purely uncorrelated noise is assumed in the photometry. These simplifications are useful for a rapid initial analysis, but will yield underestimated uncertainties. As extensively discussed in the context of previous observations of transiting planets with high-cadence photometry, the uncertainty on the density ($M\, R^{-1/3}$) of the primary and the systematics in the photometric noise are actually the dominant sources of uncertainties on most system parameters, including the planetary radius, inclination angle and transit duration. This explains why our uncertainty intervals are larger than those quoted in F09, in spite of being based on a larger ensemble of data. 

The larger uncertainties are also due to the fact that the M09, F09 and G09 light curves indicate slightly different shapes for the transit egress (M09 and G09 favour a longer egress, therefore a higher-latitude transit across a larger star).
The shape difference between the three lightcurves clearly illustrates that correlated noise dominates the error budget. The presence of correlated noise is also apparent in the \sophie\ sequence, since no parameters for the RM effect can reproduce the sudden jump shortly after the beginning of the data sequence.

For these reasons, as well as the arguments and examples presented in \citet{pon06}, the probability distributions described by the MCMC integration including correlated noise, uncertainties on all parameters, and host star properties constrained by stellar evolution models, arguably provide a better description of the actual implications of the data in terms of physical parameters and confidence intervals.

\subsection{Spin-orbit angle}

The posterior probability distribution for the projected spin-orbit angle $\lambda$ corresponds to configurations in which the planet crosses mainly the receding limb of the star. An aligned orbit ($\lambda=0$) is excluded to a $>95$\% level of confidence. The MCMC calculation shows that a spin-orbit alignment in the system is made unlikely by the combination of the shape of the SOPHIE radial velocity time series during transit and the fact that the MEarth photometry excludes a low-latitude transit by placing a strict upper limit on the transit duration. 

The result is sensitive to the level of correlated noise assumed in the radial velocity data. This is easily understood: since the amplitude of the RM anomaly is of the order of 15 m$\,$s$^{-1}$, assuming systematics of similar amplitude can reconcile the observed shape with that expected in case of spin-orbit alignment by attributing most of the observed variation to systematics. The presence of some systematics in the SOPHIE data is apparent from the mismatch between the first few measurements during transit, to the level of a few  m$\,$s$^{-1}$, but the excellent correspondence between the moment of photometric and spectroscopic egress indicates that the systematics do not dominate the SOPHIE sequence, and therefore that such a high value of $\sigma_r$ is unlikely (note that in our standard solution we already increase the correlated noise for the first measurements in the night to 12 m$\,$s$^{-1}$, which is conservative).

Therefore, our conclusion is that the combined data indicate a tilted system, with the planet crossing the receding side of the star during the transit.
\cible\ is  the second highly tilted planetary orbit known, after XO-3 \citep{heb09,win09b}, out of the twelve measured systems \citep{fab09}. More strikingly, four of these planets have eccentric orbits, and two of those have tilted orbit. We may therefore be seeing early indications that in addition to having a  broad eccentricity distribution, extrasolar gas giants have a bimodal distribution of spin-orbit angles \citep{fab09}. A possible caveat is that both XO-3 and \cible\ are high-mass gas giants, which may have formed differently than Jupiter-mass planets \citep{rib07}.

\subsection{Planetary radius}

The size of the planet is well constrained by the data, in spite of the fact that only the transit egress was observed. This may seem surprising. It is due to the observation of the secondary eclipse and the peculiarity of the orbit of HD80606b: the planet is much nearer to the star during the secondary eclipse than during the transit, because of the large eccentricity and the position of the periastron. As a result, the mere fact that the planet is transiting implies that the secondary eclipse occurs at low latitude behind the star (see Fig.~\ref{geom}). This lifts the usual degeneracy between stellar radius and impact parameter, and means that the former can be derived reliably from the duration of the secondary eclipse, even in the absence of the constraint from the fractional duration of the ingress/egress. The probability distribution of the stellar radius is further narrowed by the Bayesian prior on the inclination angle, which penalizes high-latitude transits,  by the shape of the transit egress, and by stellar evolution models.

The main remaining uncertainty on the size of the planet is due to global systematics in the photometry. The slightly longer egress duration favoured by the M09 and G09 data would indicate a higher impact parameter, therefore a larger star and a bigger planet. However, this is partly compensated by the fact that these datasets also favour a slightly shallower transit, therefore a smaller planet. Altogether, the uncertainty on the planetary radius is small enough for useful comparison with models. % (actually smaller than the difference between equatorial and polar radius for Solar-system gas giants). 

We confirm that \cible\ is not a ``bloated''  hot Jupiter, as discussed in M09, and has a size compatible with current models for irradiated gas giants. Given its large mass,  a planet like HD 80606b would require more additional energy that Jupiter-mass planets to be inflated to a higher radius. However, two other exoplanets with similar masses, CoRoT-Exo-2b ($3.3 M_J$) and OGLE2-TR-L9b ($4.5 M_J$) have radii around 1.5 $R_J$, showing that such inflated sizes are indeed possible for these heavy planets. The two other known transiting planets in the 3-5 M$_{\rm J}$ mass range, HD 17156b and WASP-10b, have radii of $1.0$ and $1.1 R_J$ respectively.

The size of these planets is conspicuously correlated with the amount of flux received from the parent star. CoRoT-Exo-2b and OGLE2-TR-L9b follow close orbits (1.7 and 2.4-day periods), while HD 17156b and HD 80606b have the widest orbits of known transiting planets, with WASP-10b being an intermediate case. The observed correlation between size and incident flux for transiting gas giants may therefore extend to several Jupiter masses, which provides another benchmark that any successful explanation of anomalous exoplanet radius must meet.

\begin{figure}
\resizebox{8cm}{!}{\includegraphics{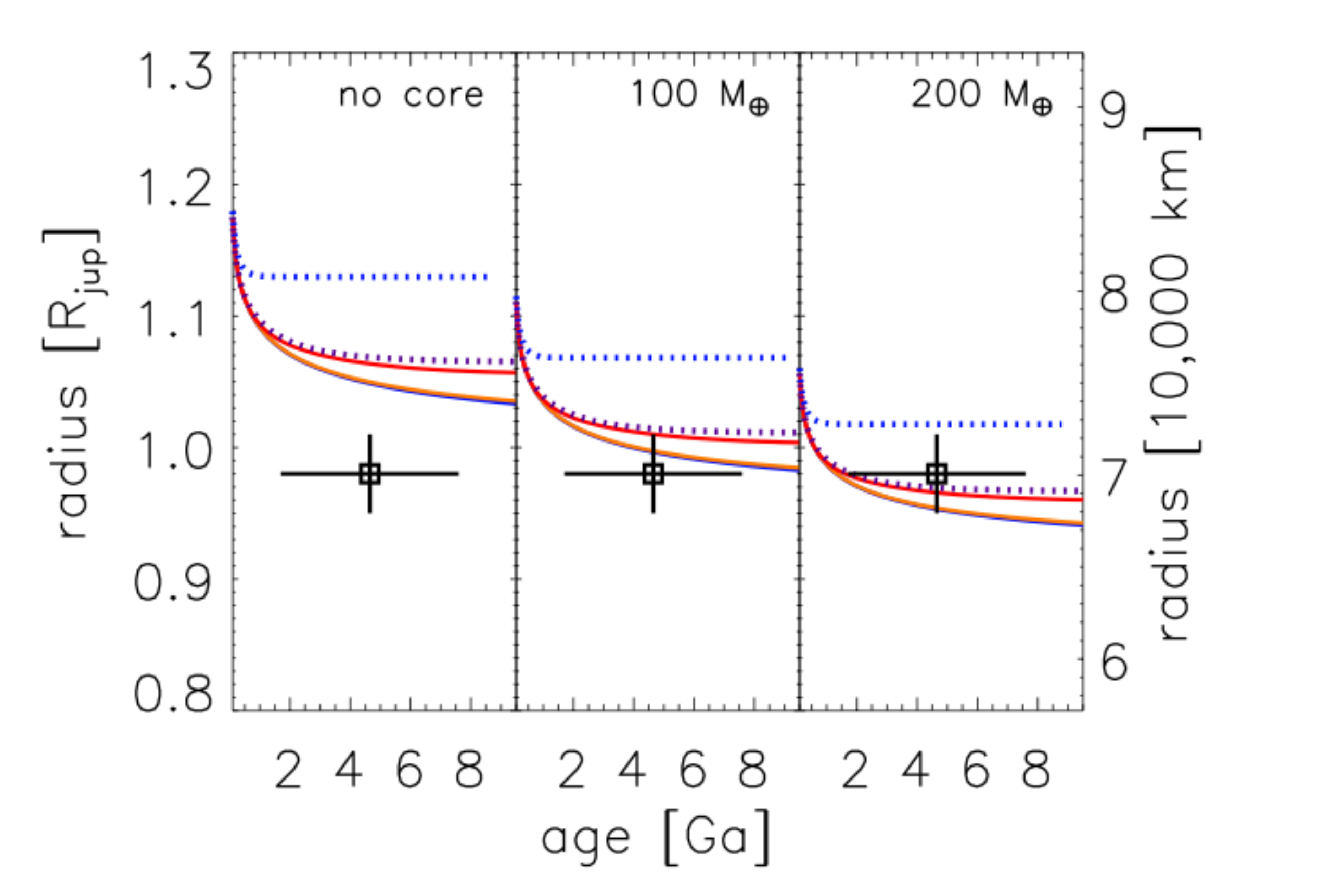}}
\caption{Evolution of the planetary radius according to models for HD 80606b. {\bf Left:} without heavy-element core. {\bf Middle:} with a 100\,$M_{\oplus}$ core. {\bf Right:} with a $200\,M_{\oplus}$ core. From bottom to top, the lines correspond to different assumption on the physics of the evolution models used to match the sizes of transiting planets: blue: standard models with no extra heating; orange: dissipating 0.5\% of incoming stellar heat flux energy in the interior of the plane ; red: opacities increased by a factor 30. The dotted (blue) lines correspond to models with an extra luminosity dissipated at the center of the planet $L_{\rm tide}=10^{26}$ and $10^{27}\rm\,erg\,s^{-1}$, respectively. }
\label{model}
\end{figure}

Figure~\ref{model} shows the evolution of the radius of HD 80606b according to models described in \citet{gui06}, with the position of our reference solution indicated. The different panels show the radius evolution of a planet with a solar-composition envelope (containing about 25\,M$_\oplus $ in heavy elements), but either without core (left panel), with a 100\,M$_\oplus$ core (middle panel) or with a 200\,M$_\oplus$ core (right panel). Clearly, although uncertainties in the distribution of heavy elements and equations of state have to be taken into account \citep[e.g.][]{bar08}, our models point towards the existence of a large mass in heavy elements (at least 60\,M$_\oplus$) present in the planet. All possibilities above 60 and 200$_\oplus$ are consistent with the measured radius. Solutions with even larger cores are possible, but we regard them as unlikely given that realistic critical core masses are smaller than 100\,M$_\oplus$ \citep{iko06} and planets with masses larger than Jupiter tend to scatter planetesimals much more efficiently than they accrete them \citep[e.g.][]{gui00}.

Given that HD80606 is metal-rich, this large inferred core mass is consistent with a correlation between heavy elements in the star and in the planet \citep{gui06,bur07,gui08}. It should be noted that if the planet had the same composition as the star, its total mass of heavy elements should be of order 50\,M$_\oplus$, clearly smaller than the estimates that are derived here.

Interestingly, using the concept of a maximum mass in heavy elements in the planet yields an upper limit on the rate of tidal dissipation that it may undergo. As shown by the dotted lines in Fig.~\ref{model}, a maximum rate of dissipation compatible with the observations and models with core of at most 200\,M$_\oplus$ is of order $10^{27}\rm\,erg\,s^{-1}$. While this may seem large, it is several times smaller that the dissipation due to tides for this planet if it has been brought in by a Kozai mechanism \citep{wu03}, estimated by $GM_* M_{\rm p}/(a_{\rm p} \tau_{\rm migration})\approx 5\times 10^{27}\,\rm erg\,s^{-1}$ (where we used $a_{\rm p}$ as the present semi-major axis of the planet, and $\tau_{\rm migration}\approx 1\,$Ga from  WM03). Several possibilities exist: (i) The planet may have a tidal Q larger than the $3\times 10^{5}$ assumed by WM03 so that it would have migrated on a longer timescale; (ii) Heat dissipation may occur in a shallow region close to the atmosphere of the planet and be reradiated quickly after the approach to the star; (iii) The planet may have a core that is much larger than envisioned in the present study. Except in the last case, our calculations indicate that the planet's intrinsic effective temperature should be of order 200 to 300\,K, much smaller than the 600 to 700\,K envisioned by \citet{wu03} and \citet{lau09}.

%During its closest approach at periastron, the upper hydrogen atmosphere of HD~80606b could be blown off by the extreme ultraviolet radiation of the G5 star. At periastron, the distance of the planet to its star is similar to that of the evaporating hot Jupiter HD~209458b. But HD~80606b spends less than 2~days (1/56 of its period) at stellar distances $< 0.1$~AU, above which atmospheric evaporation is unlikely to significantly erode the atmosphere and affect the planet life time. Consequently, the optically thick radius of the planet in the ultraviolet, which could be measured during transit using, e.g., the \emph{Hubble Space Telescope}, should not be excessively larger ) than the optically thick radius measured in the visible, unlike in the case of HD~209458b. Future observations of the primary transit in the UV could thus confront the theory of the evaporation of hot Jupiters with an a priori non-evaporating `mild Jupiter' case. 

\subsection{Future observations}

Given the brightness of the host star and its peculiar orbit, HD 80606 is likely to remain an important target for further studies. Table~\ref{ephem} shows the predicted times of ingress and egress for the next transit events according to our reference solution and 68\% central confidence intervals. 

\begin{table}[h]
\centering
\begin{tabular}{l l l}
\hline
Date 			& ingress [BJD] 	& egress [BJD]            \\
\hline
5 June 2009 		& $2454987.53 \pm 0.05$	& $2454987.974 \pm 0.005$ \\
24 September 2009 	& $2455098.96 \pm 0.05$	& $2455099.409 \pm 0.005$ \\
13-14 January 2010 	& $2455210.40 \pm 0.05$	& $2455210.844 \pm 0.006$ \\
5 May 2010	 	& $2455321.84 \pm 0.05$	& $2455322.281 \pm 0.006$ \\
24-25 August 2010 	& $2455433.27 \pm 0.05$	& $2455433.717 \pm 0.006$ \\
14 December 2010 	& $2455544.71 \pm 0.05$	& $2455545.152 \pm 0.007$ \\
4-5 April 2011 		& $2455656.14 \pm 0.05$	& $2455656.588 \pm 0.008$ \\
\hline
\end{tabular}
\caption{Predicted ephemerides for the next transit events}
\label{ephem}
\end{table}

When the transit duration will be measured by observation of future transit events, the corresponding value of the most probable impact parameter and planetary radius can be read directly off Fig.~\ref{mc_fig}.

\section{Conclusion}

Our analysis of combined photometric and spectroscopic observations of HD 80606 have led to new estimates of the stellar and planetary parameters of the system. These values allow relatively tight constraints on the planet's composition, with a likely mass in heavy elements between 60 and 200\,M$_\oplus$, which imply a rather efficient accretion of planetesimals most probably during the planet's formation and early evolution. 

We have also shown that the planet's orbit is probably not aligned with the star's spin: this strengthens the argument that the planet may have migrated inwards through a Kozai mechanism and thus acquired its large eccentricity (Wu \& Murray 2003). 

Our analysis can be readily put to the test with future observations of planetary transits, as we predict a total duration of the eclipse of $T_{1-4} = 11.9\pm 1.3$ hours. Furthermore, the analysis of the planet's primary and secondary transits, and of the planet's irradiation in the infrared when far from the star will also be extremely fruitful to understand how tidal energy is dissipated: we predict that the intrinsic effective temperature of the planet is between 200 and 300K, smaller than the zero-albedo mean equilibrium temperature due to stellar irradiation, 400K. In contrast, larger effective temperatures ($\sim$700K) have been hypothetized, due to efficient dissipation of heat in the planet (Laughlin et al. 2009).

\begin{acknowledgements}
Financial support for the \sophie\ Consortium 
from the "Programme national de plan\'etologie" (PNP) of CNRS/INSU, 
France, and from the Swiss National Science Foundation (FNSRS) 
are gratefully acknowledged. We also acknowledge support from the UK Science and Technology Facilities Council, the French
CNRS and Funda\c{c}\~ao 
para a Ci\^encia e a Tecnologia, Portugal.
\end{acknowledgements}

\bibliographystyle{aa}
\bibliography{art606_arxiv}{}

\end{document}